**Electrical and thermal transport properties of medium-entropy $Si_yGe_ySn_x$ alloys**

Duo Wang,[1] Lei Liu,[1] Mohan Chen,[2,*] and Houlong Zhuang[1,*]

[1]School for Engineering of Matter, Transport and Energy, Arizona State University, Tempe, AZ 85287, USA

[2]CAPT, HEDPS, College of Engineering, Peking University 100871, P.R. China

*mohanchen@pku.edu.cn

†zhuanghl@asu.edu

**Abstract**

Electrical and thermal transport properties of disordered materials have long been of both theoretical interest and engineering importance. As a new class of materials with an intrinsic compositional disorder, high/medium-entropy alloys (HEAs/MEAs) are being immensely studied mainly for their excellent mechanical properties. By contrast, electrical and thermal transport properties of HEAs/MEAs are less well studied. Here we investigate these two properties of silicon (Si)-germanium (Ge)-tin (Sn) MEAs, where we keep the same content of Si and Ge while increasing the content of Sn from 0 to 1/3 to tune the configurational entropy and thus the degree of compositional disorder. We predict all $Si_yGe_ySn_x$ MEAs to be semiconductors with a wide range of bandgaps from near-infrared (0.28 eV) to visible (1.11 eV) in the light spectrum. We find that the bandgaps and effective carrier masses decrease with increasing Sn content. As a result, increasing the compositional disorder in $Si_yGe_ySn_x$ MEAs enhances their electrical conductivity. For the thermal transport properties of $Si_yGe_ySn_x$ MEAs, our molecular dynamics simulations show an opposite trend in the thermal conductivity of these MEAs at room temperature, which decreases with increasing compositional disorder, owing to enhanced Anderson localization and strong phonon-phonon anharmonic interactions. The enhanced electrical conductivity and weakened thermal conductivity make $Si_yGe_ySn_x$ MEAs with high Sn content promising functional materials for thermoelectric applications. Our work demonstrates that HEAs/MEAs not only represent a new class of structural alloys but also a novel category of functional alloys with unique electrical and thermal transport properties.

**Introduction**

The Seebeck effect refers to the presence of electronic potential between two junctions in a material at different temperatures. The Peltier effect, on the other hand, illustrates the heat generation or loss at an electrified junction of two materials under voltage. These two facts provide the basis of the thermoelectric (TE) effect for the energy conversion between electrical and thermal resources. While the TE effect is universal in all materials, in most of them the effect is too trivial to be utilized [1]. A TE material needs to satisfy multiple criteria to achieve the energy conversion between heat and electricity. The maximum efficiency of the conversion is measured by the TE figure of merit depending on four parameters, i.e., $zT = \frac{S^2 \sigma}{\kappa} T$, where $S$ is the Seebeck coefficient, $\sigma$ stands for the electrical conductivity, $\kappa$ is the thermal conductivity, and $T$ denotes the absolute temperature of the operating environment [2, 3]. According to the definition of $zT$, an outstanding TE material at a specified temperature requires to possess a high Seebeck coefficient, a high electrical conductivity and a low thermal conductivity. Because of the interdependency of the four parameters, it is not always feasible to increase $zT$ by tuning one of the parameters without affecting the others. For example, the Seebeck coefficient decreases with increasing carrier density that increases the electrical conductivity [4, 5].

Efforts have been made to search for suitable TE materials with high $zT$. Metals generally have high electrical conductivities, but they also display high thermal conductivities, which balance out the $zT$ and cannot function as suitable TE materials. By contrast, semiconductors, whose $zT$ are generally much higher than metals, have found their great potential in TE applications by virtue of tunable TE properties via alloying. For example, alloying cobalt triantimonide ($CoSb_3$), with iridium (Ir) of similar electronegativity to Co, $CoSb_{3-x}Ir_x$ tends to show covalent bonding and high carrier mobility, which keeps the electrical conductivity unaffected while reducing the thermal conductivity. As a result, an optimized $zT$ of 0.8 is obtained at 600 K [5]. Lead chalcogenides alloys, $(PbTe)_{1-2x}(PbSe)_x(PbS)_x$, serve as another example of increasing the TE performance by coalloying the elements from the same group in the periodic table. The addition of PbTe and PbSe to PbS maintains the stable and homogeneous rock-salt structure and decreases the thermal conductivity of PbS by increasing the scattering strength of phonon resulted from the compositional disorder [6]. As a result of this alloying strategy, the TE performance of $(PbTe)_{1-2x}(PbSe)_x(PbS)_x$ is superior to that of individual binary Pb$X$ ($X$ = S, Se, or Te) alloys.

Although many materials have been reported to show outstanding $zT$ at high temperatures such as PbTe with $zT$ of 0.8 at 580 K [7], its $zT$ drop drastically to 0.001 at room temperature, making it unsuitable for room-temperature TE applications. At the same time, TE materials that are able to operate near room temperature are important for energy-conversion applications in different fields. For instance, the industrial waste heat common from the recycling water is always discarded into environment via the form of exhaust gas or air flow. A major portion of the heat from the low-temperature recycling water/steam is commonly wasted because of lack of commercial conversion devices [8]. As another heat resource, hot springs, one clean, cost-effective, and sustainable geothermal resource, have a temperature of 347 K [9] can be employed. Both of these examples require TE materials that can be efficient in power conversion near room temperature. Only a handful of materials are found to still display considerable $zT$ for TE applications at room temperature. For example, $CsBi_4Te_6$ consists of a $Cs_+$ layer that separates $[Bi_4Te_6]_-$ bilayers. Owing to this complex structure, the phonons in $CsBi_4Te_6$ exhibit long and convoluted phonon mean free paths that significantly reduce the thermal conductivity [10, 11]. As a result, $CsBi_4Te_6$ can reach a high $zT$ of 0.8 and a low thermal conductivity of 1.25 W/(m·K) at room temperature [11]. Among the limited number of TE candidates at room temperature, challenges in manufacturing persist, hindering their wide industrial applications in TE devices. Specifically, fabricating the room-temperature TE materials such as $CsBi_4Te_6$ requires an intensive control of the synthesis process. Therefore, it is important to identify room-temperature TE materials that are amenable to modern manufacturing technologies.

High/medium-entropy alloys (HEAs/MEAs; Following the convention [12, 13], if the number of elements is greater than or equal to five, HEAs is used; otherwise, MEAs is a more accurate terminology.) recently emerge as alternative candidates for TE applications. Different from conventional alloys that have one principal element and the other elements are added to tune the properties [14, 15], HEAs/MEAs represent a novel category of alloys with multiple principal components, where all the constituent elements have equal or comparative concentrations [16]. HEAs/MEAs generally exhibit three different phases: solid solution, intermetallic compound, or the mix of these two phases [17, 18]. In comparison with conventional alloys, the disorder atomic arrangement in HEAs/MEAs is associated with larger configurational entropy. Including the entropic contribution to the energy, HEAs/MEAs could form stable solid solutions at higher temperatures [19, 20]. Many HEAs/MEAs exhibit excellent mechanical properties such as elevated-temperature strength [21, 22] and high elastic modulus [23-25]. In addition to mechanical properties, HEAs/MEAs display outstanding functional properties in the applications from superconductors [26-28] to semiconductors with adjustable bandgaps [16]. HEAs/MEAs can be suitable TE materials mainly for two reasons. First, single-phase HEAs/MEAs can maintain a stable solid solution phase with high crystalline symmetry, across which electrons move freely. In line with the concept of 'phonon-glass electron-single-crystal' [29, 30], single-phase HEAs/MEAs are expected to show difficulty in conducting phonons but easiness in transporting electrons. Second, the thermal conductivity of HEAs/MEAs can be suppressed by several phonon scattering mechanisms that involve point defects, lattice distortions, and compositional disorder, which are commonly seen in HEAs/MEAs [31, 32]. Indeed, HEAs/MEAs such as Al$_x$CoCrFeNi [30], PbSnTeSe [33], and BiSbTe$_{1.5}$Se$_{1.5}$ [31] have been synthesized and employed as TE materials suitable for high (>1000 K), medium (500-900 K), and low working temperatures (300-500 K), respectively.

Manufacturing techniques of Group IV alloys consisting of silicon (Si), germanium (Ge), tin (Sn) are supported by the sophisticated semiconductor industry, it is therefore worthwhile exploring Si-Ge-Sn alloys for TE applications. Binary Si-Ge alloys have long been used as TE materials in waste heat generation and conversion for automotive applications [34-36]. In particular, binary Si-Ge alloys exhibit excellent TE performance at high temperature (> 1000 K) [37-39]. Experiments have also shown that adding Sn to Si-Ge alloys further lowers the thermal conductivity without significantly affecting the electrical conductivity [40]. Consistent with experiment, theoretical studies have found that the thermal conductivity of Si-Ge-Sn alloys can be reduced through controlling the Sn content and decreasing the bulk to thin films [41]. Comparing to Si or Ge atoms, a Sn atom is heavier and has a much larger radius. Heavier and larger Sn atoms in Si-Ge alloys increases the compositional disorder and causes a lattice distortion (see below), leading to strong anharmonic phonon-phonon scattering and thereby decreasing the thermal conductivity [42]. Si-Ge-Sn alloys exhibit a wide range of band gaps up to 1.4 eV that vary with composition and the bandgap types switch between direct and indirect [40, 43-45]. According to the empirical "10$k_B T$" rule proposed by Mahan [46, 47], a TE material operating at temperature $T$ should have a band gap of around 10$k_B T$ to maximize the $zT$. Composition-dependent bandgaps in Si-Ge-Sn alloys therefore enable them to achieve optimal TE performance at various operating temperatures [41, 48].

In this work, instead of exploring the regions in the composition space representing conventional Si-Ge-Sn alloys that have been the subject of a number of theoretical and experimental studies [14, 41, 49, 50], we focus on a special region of the composition space (see Figure 1(a)), where we keep the concentration of Si and Ge equal while varying the concentration of Sn from 0 to 1/3. The chemical formula of these special Si-Ge-Sn alloys can therefore be written as $Si_yGe_ySn_x$, where $2y + x = 1$ and $x \leq 1/3$. In choosing these concentrations, we are able to obtain the trends on how the electrical and thermal conductivities and the TE figure of merit are modified via systematically increasing the configuration entropy and therefore the extent of compositional disorder. Figure 1(b) shows that the configurational entropies of $Si_yGe_ySn_x$ MEAs calculated from the Boltzmann entropy formula increase from 0.71 $k_B$ ($k_B$ : the Boltzmann constant) at $x = 0$ to 1.10 $k_B$ at $x = 1/3$. We have recently studied the $Si_{1/3}Ge_{1/3}Sn_{1/3}$ MEA, which exhibits a single phase without a short-range chemical order, a near-infrared bandgap and a wide range of vacancy formation energies [16]. Although no experiments have yet been performed to fabricate this special series of $Si_yGe_ySn_x$ MEAs and it is also challenging to fabricate the Si-Ge-Sn alloys with high Sn content because of large lattice mismatch [16, 41, 51-53], recent experimental techniques developed by Kouvetakis and coworkers to fabricate conventional Si-Ge-Sn alloys with high Sn content may be adjusted to synthesize these MEAs [53]. Before the experiment becomes available, we here predict the structural, electrical, and thermal transport properties of $Si_yGe_ySn_x$ MEAs and explore their TE applications.

**Simulation Methods**

To create simulation cells for the density functional theory (DFT) [54, 55] calculations of $Si_yGe_ySn_x$ MEAs, we first generate a $3 \times 3 \times 3$ supercell from an 8-atom unit cell of cubic Si with the diamond structure. The supercell thus consists of 216 Si atoms and we randomly permutate the locations of these atoms and substitute a portion of the atoms with Ge and Sn atoms, depending on the Sn content $x$. Given the fixed total number of atoms in the supercell, we obtain 11 structures of $Si_yGe_ySn_x$ MEAs with different concentrations of Sn, namely, $x = 0$, 1/108, 1/54, 1/36, 1/27, 1/18, 1/12, 1/9, 1/6, 1/4, and 1/3. Figure 1(a) illustrates the structure of a $Si_yGe_ySn_x$ MEA with the highest configurational entropy (i.e., $x = 1/3$).

We use the Vienna Ab initio Simulation Package (VASP; version 5.4.4) for all the DFT calculations and the Perdew-Burke-Ernzerhof (PBE) [56] functional to describe the exchange-correlation interactions in geometry optimizations and energy calculations. To remedy the problem of bandgap underestimation due to the PBE functional [57], we apply the modified Becke-Johnson (mBJ) exchange potential, which has been shown to significantly improve the bandgaps of many semiconductors such as aluminum phosphide (AlP), silicon carbide (SiC), and gallium arsenide (GaAs) [58]. We compute the electronic structures of $Si_yGe_ySn_x$ using this potential based on the fully optimized structures with the PBE functional. The mixing parameter $c$ in the potential is determined self-consistently. To describe the electron-nuclei interactions, we use the standard PBE version of Si, Ge, and Sn potential datasets generated via the projector augmented-wave (PAW) method [59, 60] In these PAW potentials, the $3s^2$ and $3p^2$ electrons of Si atoms, the $4s^2$ and $4p^2$ electrons of Ge atoms, and the $5s^2$ and $5p^2$ electrons of Sn atoms are adopted as valence electrons. The plane wave cutoff energy is 400 eV. We fully optimize the 216-atom supercells using a $2 \times 2 \times 2$ Monkhorst-Pack $k$-point grid [61] and the force convergence criterion during the geometry optimization calculations is set to 0.01 eV/Å.

To calculate the Seebeck coefficient and electrical conductivity of $Si_yGe_ySn_x$ MEAs, we use the BoltzTraP2 package [62], which solves the linearized electron Boltzmann transport equation according to rigid-band approximation (rather than the constant relaxation time approximation as implemented in the BoltzTraP package [63]) assuming that the band structure is not affected by temperature and doping. The doping and temperature effects are taken into account in the Fermi-Dirac distribution function $f(\varepsilon,\mu,T)$. The inputs for our BoltzTraP2 calculations are the band energies $\varepsilon$ obtained from the mBJ exchange potential and the key outputs are the Seebeck coefficient and electrical conductivity tensors that depend on the electron chemical potential $\mu$ from the following equations [62]:

$$S = \frac{1}{qT} \frac{\int \sigma(\mu,T)(\varepsilon-\mu)\left(-\frac{\partial f(\varepsilon,\mu,T)}{\partial \varepsilon}\right) d\varepsilon}{\int \sigma(\mu,T)\left(-\frac{\partial f(\varepsilon,\mu,T)}{\partial \varepsilon}\right) d\varepsilon} \qquad (1)$$

and

$$\sigma = \int \sigma(\mu,T)\left(-\frac{\partial f(\varepsilon,\mu,T)}{\partial \varepsilon}\right) d\varepsilon, \qquad (2)$$

where $q$ is the carrier charge and $\sigma(\mu,T)$ is called the transport distribution function according to the linearized electron Boltzmann transport equation. $\sigma(\mu,T)$ describes the carrier transport property that depends on the group velocity, concentration, and lifetime of the carriers. We are not considering the anisotropy in this work, so we report the average Seebeck coefficient and electrical conductivity in the three dimensions. To compute the electrical conductivity, we use a constant relaxation time of 10 femtoseconds, which has been widely used in computing the electrical conductivities of many intrinsic or lightly doped semiconductors [63-66].

To obtain the thermal conductivity of Si$_y$Ge$_y$Sn$_x$ MEAs, we perform classical equilibrium molecular dynamics (EMD) simulations using the Large-scale Atomic/Molecular Massively Parallel Simulator (LAMMPS) [67]. The many-body interatomic interactions are described using a modified Stillinger-Weber potential [68] that was fitted from DFT to reproduce properties such as the phase stability and elastic constants [69, 70], and applied to study the thermal conductivity of Si-Ge-Sn alloys [42, 71]. The simulation cells used in MD simulations are much larger than those used in the DFT calculations, and the way we create simulation cells is similar. That is, the MD supercells are based on a $9 \times 9 \times 9$ supercell of the Si unit cell and each supercell consists of 5832 atoms. Because of thermal fluctuations, we create three sets (using three different random seeds) of 5832-atom supercells for each Si$_y$Ge$_y$Sn$_x$ MEA and report the average thermal conductivity. The thermal conductivity of Si$_y$Ge$_y$Sn$_x$ MEAs at room temperature is calculated through EMD simulations based on the Green-Kubo method [72-74]. The time step for the MD simulations is 1.0 fs. In the MD simulations, we first equilibrate the system for $10^6$ steps using the NVT canonical ensemble and the Nosé-Hoover thermostat [67, 75] and then runs another $4 \times 10^6$ steps for collecting the heat flux data every $1.25 \times 10^4$ steps in the NVE micro-canonical ensemble. The thermal conductivity $\kappa_\alpha$ in the $\alpha$ direction is calculated with the following equation [41]:

$$\kappa_\alpha = \frac{V}{k_B T^2} \int_0^\infty \langle J_\alpha(0) J_\alpha(t) \rangle dt \qquad (3)$$

where $J_\alpha(t)$ is the heat current at time $t$; $V$ and $T$ denote volume and temperature, respectively.

We also study the thermal conductivities of Si$_y$Ge$_y$Sn$_x$ MEAs at the DFT level using the Phonopy and Phono3py programs [76, 77]. Phonopy and Phono3py programs, post-process the second-order and third-order force constants, respectively, to yield the phonon frequencies, the thermal conductivity and the phonon linewidth by solving the linearized phonon Boltzmann transport equation. To obtain the harmonic phonon frequencies and anharmonic phonon-phonon interactions, we calculate the second-order and the third-order force constants using the finite-displacement method with a displacement of 0.03 Å. The second-order and third-order force constants are calculated using $4 \times 4 \times 4$ and $2 \times 2 \times 2$ supercells, respectively. The energies and forces of these geometries are evaluated via DFT calculations. It is nevertheless challenging to calculate the third-order force constant for a disordered alloy with a low symmetry, so we create the models of two artificial MEAs, $Si_{3/8}Ge_{3/8}Sn_{1/4}$ and $Si_{1/2}Ge_{3/8}Sn_{1/8}$, which, respectively, have three/four Si atoms, three/three Ge atoms, and two/one Sn atoms in their unit cells. The Sn content of the artificial $Si_{3/8}Ge_{3/8}Sn_{1/4}$ MEA is the same as one of the 11 selected $Si_yGe_ySn_x$ MEAs that has $x$ of 1/4. The total number of supercells to obtain the second-order and third-order force constants via DFT calculations is 18 and 4746, respectively. Although $Si_{1/2}Ge_{3/8}Sn_{1/8}$ is not one of the 11 $Si_yGe_ySn_x$ MEAs, its composition can be reflected by the same small unit cell with the Sn content smaller than 1/3. Correspondingly, the total number of supercells to obtain the second-order and third-order force constants via DFT calculations is 5 and 833, respectively.

**Results and Discussion**

We first investigate the structural properties of the 11 $Si_yGe_ySn_x$ MEAs. Because the shapes of all the supercells slightly deviate from cubic after geometry optimizations, we report the average length of the three lattice vectors. Figure 1(d) displays the variation of the average lattice constant $a_{avg}$ of $Si_yGe_ySn_x$ with the Sn content. We can fit the $a_{avg}$ data to a linear equation ($a_{avg} = 5.61 + 1.04x$) without introducing the bowing factor. Namely, the average lattice constant follows the Vegard law. This trend is in contrast to group IV binary alloys, which have been shown to exhibit sizable bowing factors [78, 79].

We next compute the formation energy $E_f$ of $Si_yGe_ySn_x$ HEAs, defined as the energy change of the following reaction:

$$y\text{Si} + y\text{Ge} + x\text{Sn} \rightarrow Si_yGe_ySn_x. \tag{1}$$

For a better comparison, we calculate the energies of the 216-atom supercells of Si, Ge, and $\alpha$-Sn with the diamond structure using the same number of atoms. Positive $E_f$ from the ground-state calculations indicates that the product (a Si$_y$Ge$_y$Sn$_x$ MEA) is less stable than the reactants (Si, Ge, and $\alpha$-Sn). Figure 1(e) shows that the energy changes of all the reactions are endothermic and that $E_f$ increases with the Sn concentration although the magnitude of increase is smaller at high Sn content.

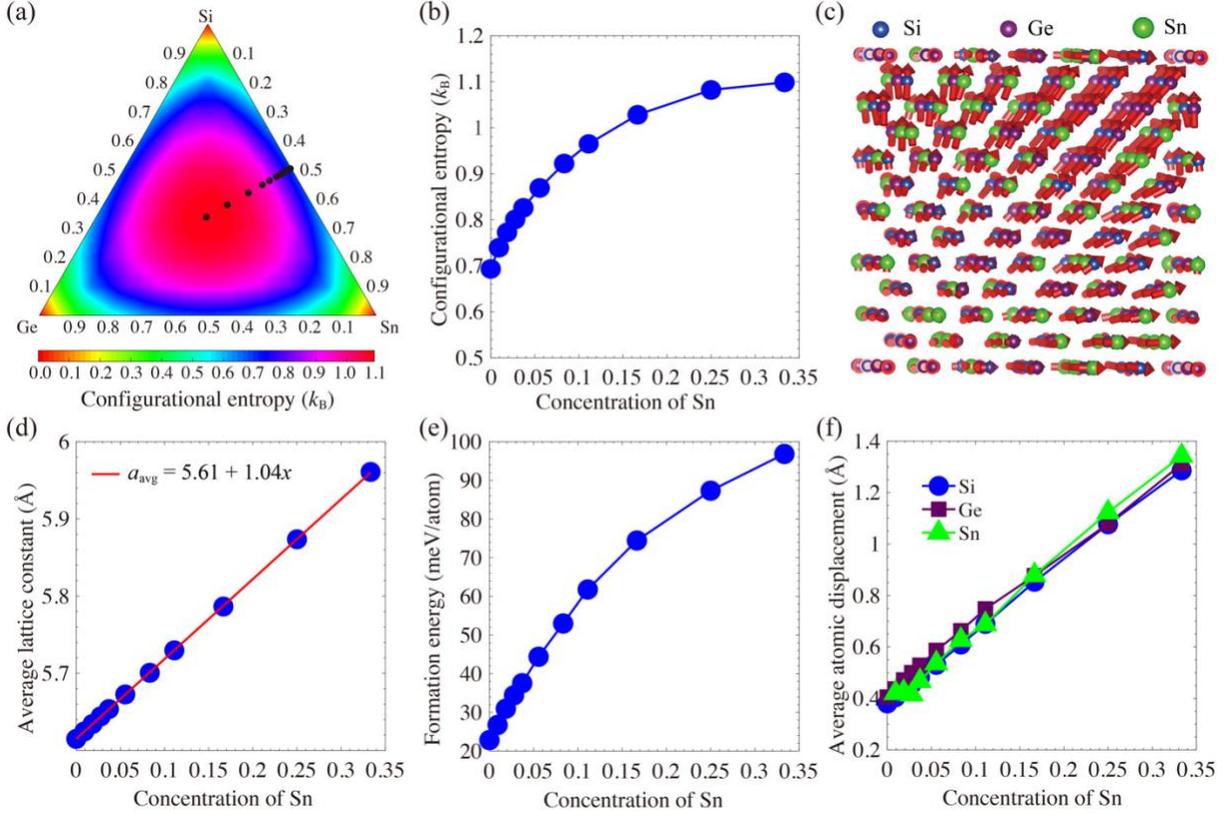

**Figure 1.** (a) Composition space of Si-Ge-Sn alloys color-coded by the configurational entropy. The black dots denote the compositions of the 11 alloys studied in this work. (b) Configuration entropy of Si$_y$Ge$_y$Sn$_x$ MEAs a function of the Sn content $x$. (c) Atomic structure of Si$_y$Ge$_y$Sn$_x$ ($x = 1/3$) MEA and displacement field showing a large lattice distortion effect. (d) Average lattice constant and its fit to a linear equation of $x$, (e) formation energy, and (f) average atomic displacement of Si$_y$Ge$_y$Sn$_x$ MEAs as a function of $x$.

We note that the configurational entropy is absent in the formation energy calculations. We estimate the entropy contribution, for example, at 500 K that is an estimated temperature at which conventional Si-Ge-Sn alloys are grown using the CVD method [40]. The configurational entropy contribution -$TS_{conf}$ to the Gibbs free energy at this temperature is about -53.3 meV/atom for the equimolar Si$_{1/3}$Ge$_{1/3}$Sn$_{1/3}$ MEA with the $E_f$ of 96.8 meV/atom. However, the Gibbs free energy remains positive (43.5 meV/atom) even considering this part of energy contribution, implying that Si$_y$Ge$_y$Sn$_x$ MEAs are out of thermodynamic equilibrium. As a result, fabricating Si$_y$Ge$_y$Sn$_x$ MEAs is likely to require the same non-equilibrium growth processes used for obtaining the conventional high-Sn content Si-Ge-Sn alloys [14, 80].

HEAs/MEAs generally exhibit a notable lattice distortion effect. To illustrate this effect, we compute the atomic displacement field for the atoms in the simulation supercell using the Si$_{1/3}$Ge$_{1/3}$Sn$_{1/3}$ MEA as an example. Figure 1(c) depicts the atomic displacement field of this MEA with the atomic positions in lattice sites of bulk Si being the reference. Each arrow represents the direction of a displacement and the length of an arrow represents the magnitude of the atomic displacement. Although the atomic displacements occur in different directions, we observe a dominant direction parallel to the (110) plane, where the atoms prefer to relocate. This direction preference causes a large lattice strain along that direction. We calculate the average atomic displacements for each element in the Si$_{1/3}$Ge$_{1/3}$Sn$_{1/3}$ MEA. As shown in Figure 1(f), the average atomic displacements for all the species increase when the Sn content is increased. This trend is caused by the fact that increasing the Sn content simultaneously creates space for Si and Ge atoms to adjust their locations. We also observe from Figure 1(f) that the average atomic displacement for the three elements is nearly the same, regardless of their different atomic radii.

Having studied the structural properties of Si$_y$Ge$_y$Sn$_x$ MEAs, we set out to understand their electronic structures and associated electrical transport properties. Figure 2 displays the band structures of the four Si$_y$Ge$_y$Sn$_x$ ($x$ = 0, 1/12, 1/6, and 1/3) MEAs calculated using the mBJ exchange potential. Due to the presence of heavy Sn atoms, we account for the spin-orbit coupling in computing the band structures. The modified Becke-Johnson exchange potential that yields accurate band gaps comparable to our recent calculation using the HSE06 hybrid density functional [81]. For example, the mBJ band gap of $Si_{1/3}Ge_{1/3}Sn_{1/3}$ is 0.28 eV, close to the HSE06 band gap of 0.38 eV [16]. As can be seen from Figure 2, the bandgap type becomes direct as the Sn content is increased to near 1/3. Figure 3(a) shows the variation of the bandgaps with the concentration of Sn, revealing that the bandgaps decrease with increasing Sn content and the range of the bandgaps, 0.28-1.11 eV, spans from visible to near-infrared in the light spectrum. Such a wide range of band gaps endow $Si_yGe_ySn_x$ MEAs with potential for applications not only in TE but also in optoelectronics devices.

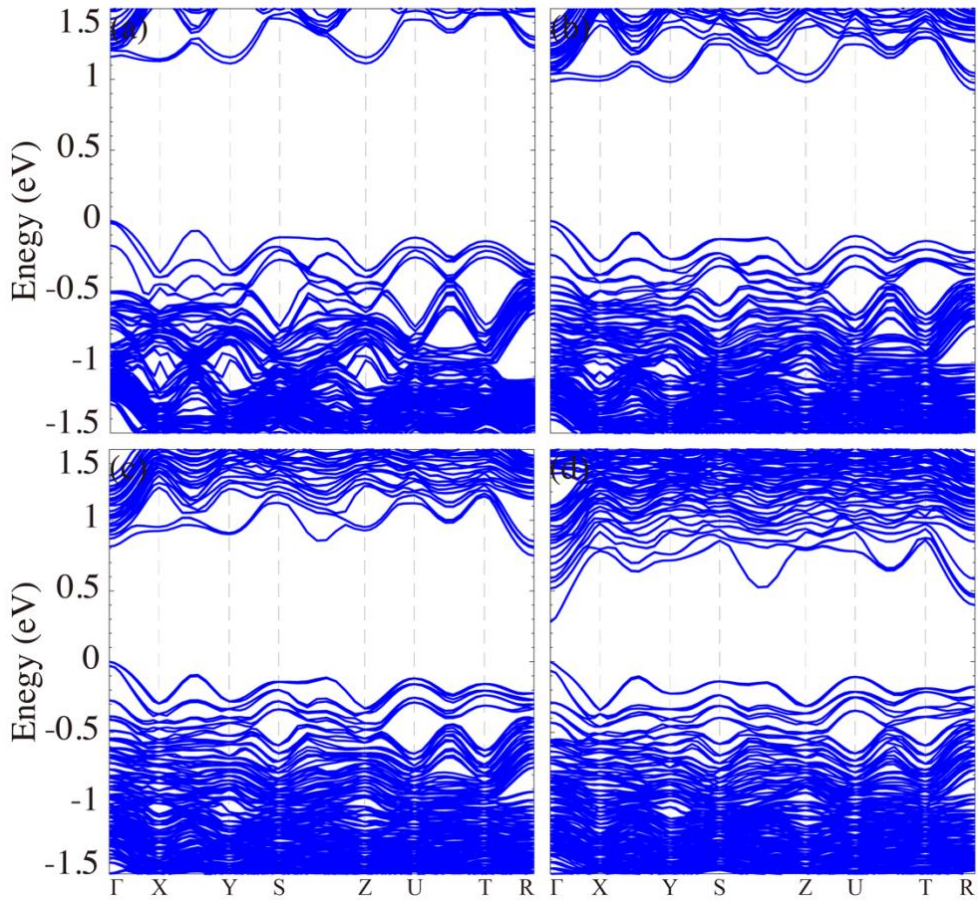

**Figure 2.** Band structures of $Si_yGe_ySn_x$ MEAs with (a) $x = 0$, (b) $x = 1/12$, (c) $x = 1/6$, and (d) $x = 1/3$. The band structures are obtained from the modified Becke-Johnson exchange potential using the optimized structures from DFT calculations with the PBE functional. Spin-orbit coupling is taken into account and the valence band maxima are set to zero. The coordinates of X, Y, S, Z, U, T, and R are (1/2,0,0), (0,1/2,0), (1/2,1/2,0), (0,0,1/2), (1/2,0,1/2), (1/2,1/2,0), and (1/2,1/2,1/2) respectively.

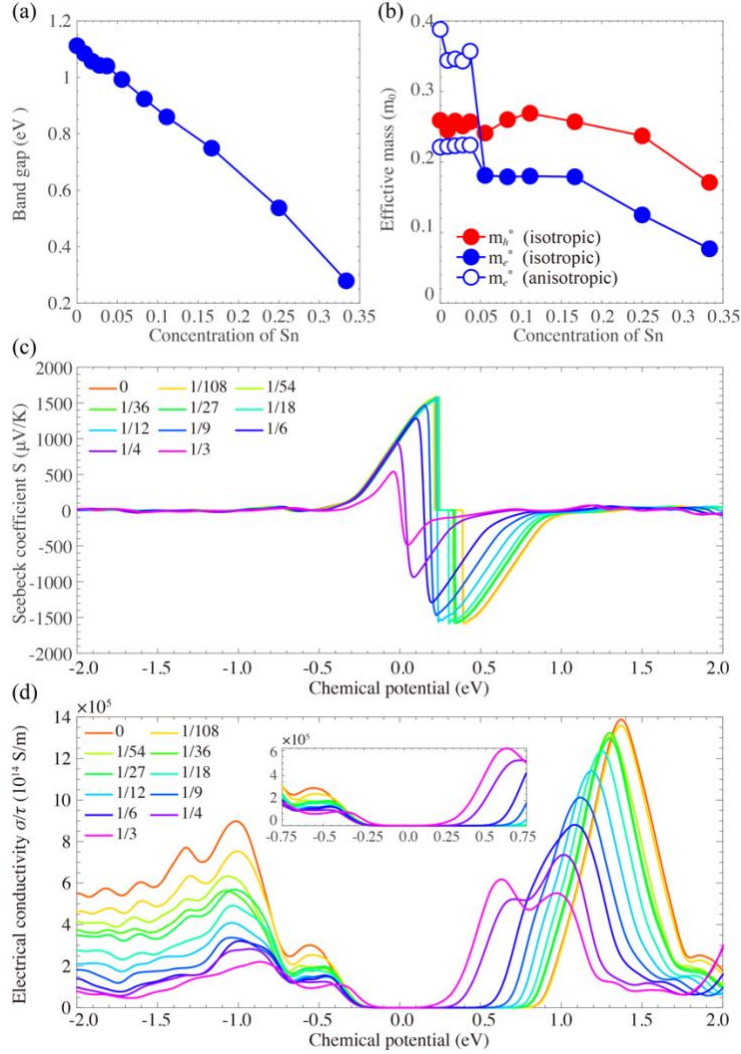

**Figure 3.** Variation of (a) bandgaps and (b) electron and hole effective masses of Si$_y$Ge$_y$Sn$_x$ MEAs with Sn content. (c) Seebeck coefficient and (d) electrical conductivity of Si$_y$Ge$_y$Sn$_x$ MEAs at different electron chemical potentials.

Figure 3(b) shows the computed electron and hole effective masses of Si$_y$Ge$_y$Sn$_x$ MEAs. The hole effective mass $m_h^*$ in the whole range of Sn content of Si$_y$Ge$_y$Sn$_x$ MEAs is isotropic, so only one set of data are shown. By contrast, the effective electron mass $m_e^*$ is anisotropic at low Sn content. Increasing the Sn content removes the anisotropy of $m_e^*$. Overall, Si$_y$Ge$_y$Sn$_x$ MEAs with high-content Sn has small effective carrier masses. In particular, Si$_{1/3}$Ge$_{1/3}$Sn$_{1/3}$ exhibits the lowest $m_h^*$ (0.171 $m_0$) and $m_e^*$ (0.077 $m_0$) among the 11 Si$_y$Ge$_y$Sn$_x$ MEAs. The low $m_h^*$ and $m_e^*$ of Si$_y$Ge$_y$Sn$_x$ MEAs with high-content Sn are comparable or even smaller than the corresponding $m_h^*$ and $m_e^*$ of Si, Ge, and GaAs [82], implying high carrier mobility in these MEAs given the same relaxation time and applied electric field.

To evaluate the room-temperature TE figure of merit of the 11 $Si_yGe_ySn_x$ MEAs, we first calculate their Seebeck coefficients and electrical conductivities at 300 K. According to Eq.1 and (2), both parameters vary with the chemical potential that can be tuned by *p/n*-type doping for a semiconductor. Figure 3(c) depicts the calculated Seebeck coefficients of the MEAs as a function of the chemical potential. We observe two typical peaks in the plots of Seebeck coefficients for each MEA, corresponding to the maximum Seebeck coefficients of the cases of *p*-type and *n*-type doping, respectively. Because of the Fermi distribution, the Seebeck coefficients decay rapidly to zero when the chemical potential significantly deviates from the band energies. For all the 11 $Si_yGe_ySn_x$ MEAs, we obtain their maximum Seebeck coefficients that range from 537 to 1583 µV/K. The maximum Seebeck coefficients of $Si_yGe_ySn_x$ MEAs with the Sn content below 1/9 have higher values ( > 1500 µV/K). Although the maximum Seebeck coefficient of $Si_{1/3}Ge_{1/3}Sn_{1/3}$ is not the largest among all the MEAs, the value of 537 µV/K appears higher than many other TE materials such as $CuGaTe_2$ and MgAgSb with the maximum Seebeck coefficients of 277 µV/K [83] and 400 µV/K [84] respectively, calculated with the PBE functional and at the same temperature.

Figure 3(d) shows the electrical conductivities (divided by the constant relaxation time $\tau = 10^{-14}$ s) of the $Si_yGe_ySn_x$ MEAs as a function of the chemical potential. We can see the ranges of chemical potentials, which match the band gaps of the 11 MEAs as shown in Figure 3(a), correspond to an electrical conductivity of zero. Outside these ranges, the electrical conductivity increases significantly. In a small potential range from -0.3 to 0.75 eV, the electrical conductivity of $Si_{1/3}Ge_{1/3}Sn_{1/3}$ is the highest among all the MEAs, where it peaks at 0.63 eV with an electrical conductivity of $6.2 \times 10^5$ S/m. For reference, the electrical conductivity is higher than many other proposed TE materials such as CuGaTe$_2$ ($1.3 \times 10^4$ S/m) [83] and Mg$_2$Si ($7.1 \times 10^4$ S/m) [85] at room temperature.

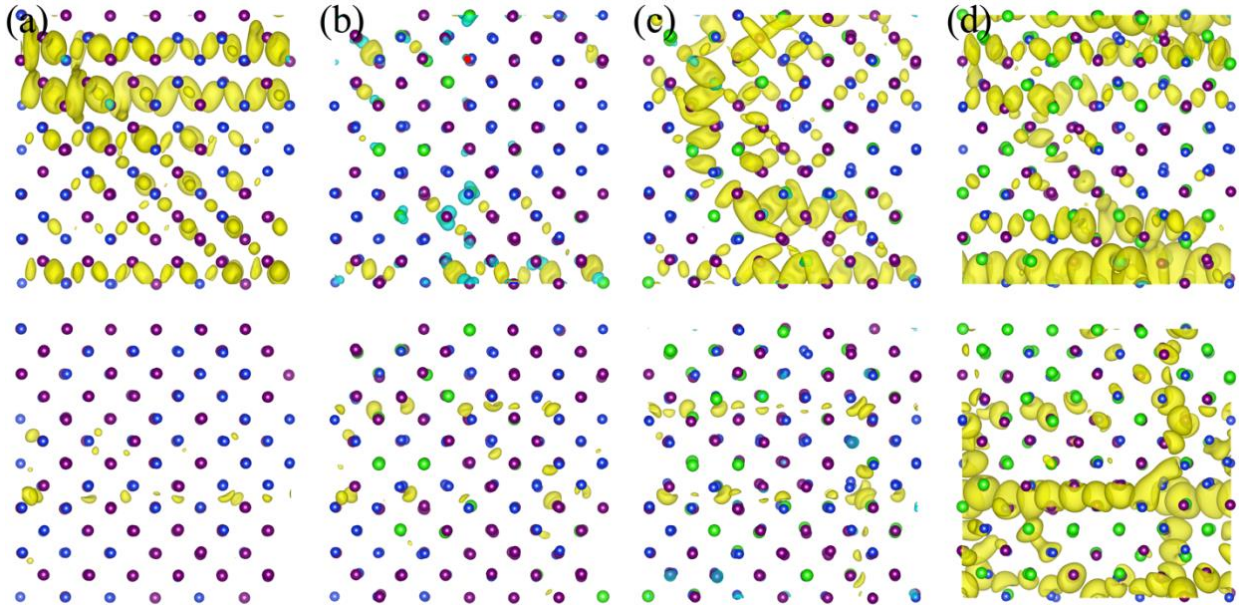

**Figure 4.** Charge density of the valence (top) and conduction (bottom) bands of Si$_y$Ge$_y$Sn$_x$ MEAs with (a) $x = 0$, (b) $x = 1/12$, (c) $x = 1/6$, and (d) $x = 1/3$. The isosurface value of Si$_y$Ge$_y$Sn$_x$ MEAs is $1.0 \times 10^{-4}$ $e$/Bohr$^3$.

The high electrical conductivity of Si$_y$Ge$_y$Sn$_x$ MEAs especially the ones with high Sn content is somewhat surprising. According to Anderson [86], one expects electron localization to occur and diminish the electrical conductivity in disordered systems like Si$_y$Ge$_y$Sn$_x$ MEAs. On the contrary, the charge densities of the top valence and bottom conductions bands of four Si$_y$Ge$_y$Sn$_x$ MEAs ($x = 0$, 1/12, 1/6, and 1/3) shown in Figure 4 reveal that, although there is carrier localization to some extent, both electrons and holes become more delocalized as the Sn content is closer to 1/3. The absence of electron/hole Anderson localization is possibly because Si, Ge, and Sn belong to the same group and have nearly the same electronegativity. As a result, the scattering between an electron and any of the three types of nuclei is expected to be similar.

We now compute the thermal conductivity of Si$_y$Ge$_y$Sn$_x$ MEAs and examine the effects of compositional disorder. Figure 5(a) shows the calculated thermal conductivity of the 11 Si$_y$Ge$_y$Sn$_x$ MEAs from the EMD simulations. The thermal conductivities of several of these MEAs have been reported in the literature using a variety of methods such as the non-equilibrium MD (NEMD) method with fixed thermostat at the end points, the reversed NEMD Müller-Plathe method, and the EMD Green-Kubo method [72, 73, 87, 88]. Our calculated thermal conductivities are consistent with the literature. For example, the thermal conductivity of Si$_{0.5}$Ge$_{0.5}$ in the present work is 1.62 W/(m·K), comparable to 1.06 ± 0.16 W/(m·K) using the NEMD method with the same interatomic potential [89, 90]. Because the thermal conductivity at the MD level of theory depends on simulation methods such as the Green-Kubo method, the simulation cell size, and most importantly, the transferability of the Si-Ge-Sn interatomic potential, we emphasize that the absolute values of thermal conductivities of Si$_y$Ge$_y$Sn$_x$ MEAs are not as important as the trend shown in Figure 5(a) that the thermal conductivity decreases with the increasing Sn content. As a matter of fact, we benchmark the thermal conductivity of Si using the same set of simulation parameters and obtain the thermal conductivity of 56.1 W/(m·K), which is only about half of the experimental value of 125.5 W/(m·K) [91]. Therefore, although our calculated thermal conductivities await experimental confirmation, they tend to be underestimated in comparison with the future experimental data.

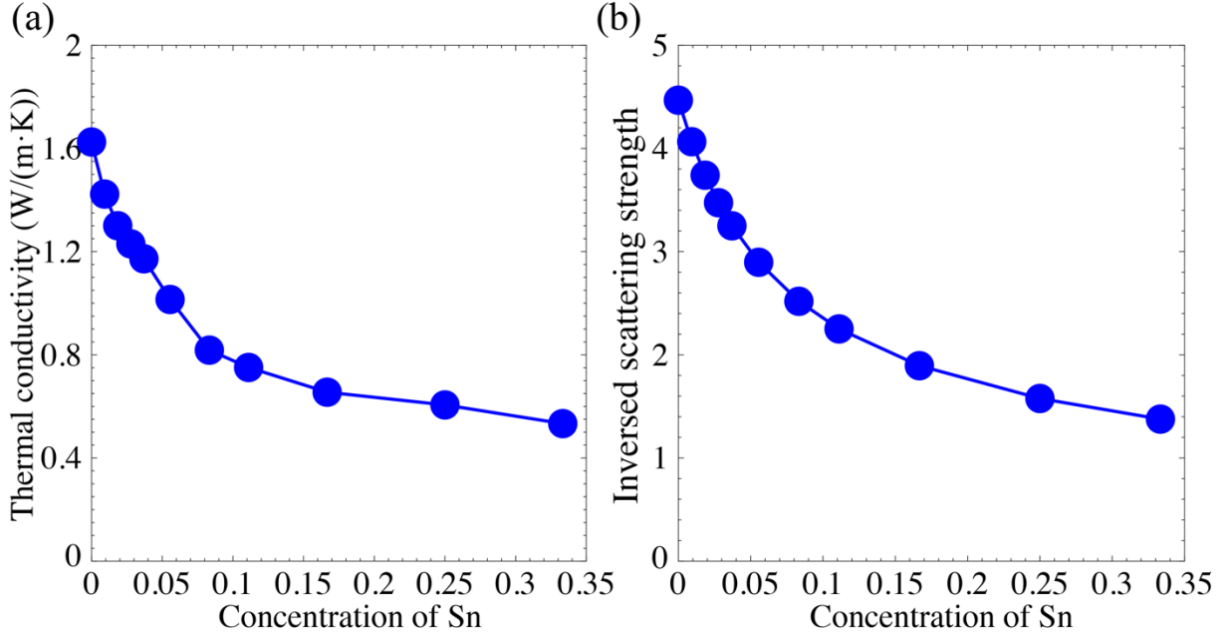

**Figure 5.** (a) Thermal conductivity of Si$_y$Ge$_y$Sn$_x$ MEAs obtained from classical molecular dynamics simulations and (b) the corresponding inversed scattering strengths calculated from Eq. 6.

Because Si$_y$Ge$_y$Sn$_x$ MEAs are semiconductors with sizable bandgaps, we do not consider the electron contribution to the thermal conductivity and therefore focus on the phonon contributions to the thermal conductivities of Si$_y$Ge$_y$Sn$_x$ MEAs. To understand the above trend of the decreased thermal conductivity of Si$_y$Ge$_y$Sn$_x$ MEAs with the increasing compositional disorder, we adopt three different perspectives related to phonon behavior: the phonon version of Anderson localization, phenomenal model of phonon scattering, and DFT calculations of phonon scattering. We start with computing the phonon vibrational frequencies and their corresponding normal modes (eigenvectors). We compute the participation ratio (*PR*) for each vibrational mode *n* using the following equation [42, 92]:

$$PR_n = \frac{\left(\sum_i e_{i,n}^2\right)^2}{N \sum_i e_{i,n}^4} \qquad (2)$$

where $e_{i,n}$ is the phonon eigenvector of mode $n$ and $N$ is the total number of atoms. Figure 6 shows the phonon density of states (PDOS) of four $Si_yGe_ySn_x$ ($x = 0$, 1/12, 1/6, and 1/3) MEAs and their corresponding $PR$ plots. We can see that the PDOS of the four MEAs appear similar except that an additional peak occurs in the $Si_{1/3}Ge_{1/3}Sn_{1/3}$ MEA near the frequency of 170 cm$^{-1}$. Another phenomenon in common is that the $PR$ values for all the phonon modes of each MEA are separated into two groups, one with low $PR$, implying large eigenvectors; the other with high $PR$, suggesting small eigenvectors. The boundary between the two groups of $PR$s is called the mobility edge. According to Allen et al. [93], the phonon modes below the mobility edge are classified as locons caused by the phonon version of Anderson localization. The mobility edge separates locons from the so-called propagons and diffusons that positively contribute to the thermal conductivity. We observe that the mobility edges of the four $Si_yGe_ySn_x$ MEAs are located at 243, 213, 179, and 122 cm$^{-1}$, respectively. Furthermore, as the Sn content increases, the $PR$ values of the locons decrease, suggesting increasingly enhanced phonon localizations, which are also manifested by the increasingly localized eigenvectors of the phonon modes with small $PR$s as displayed in Figure 7. The more and more localized phonons with the Sn content therefore explains the observed trend in the thermal conductivities (see Figure 5(a)) of $Si_yGe_ySn_x$ MEAs.

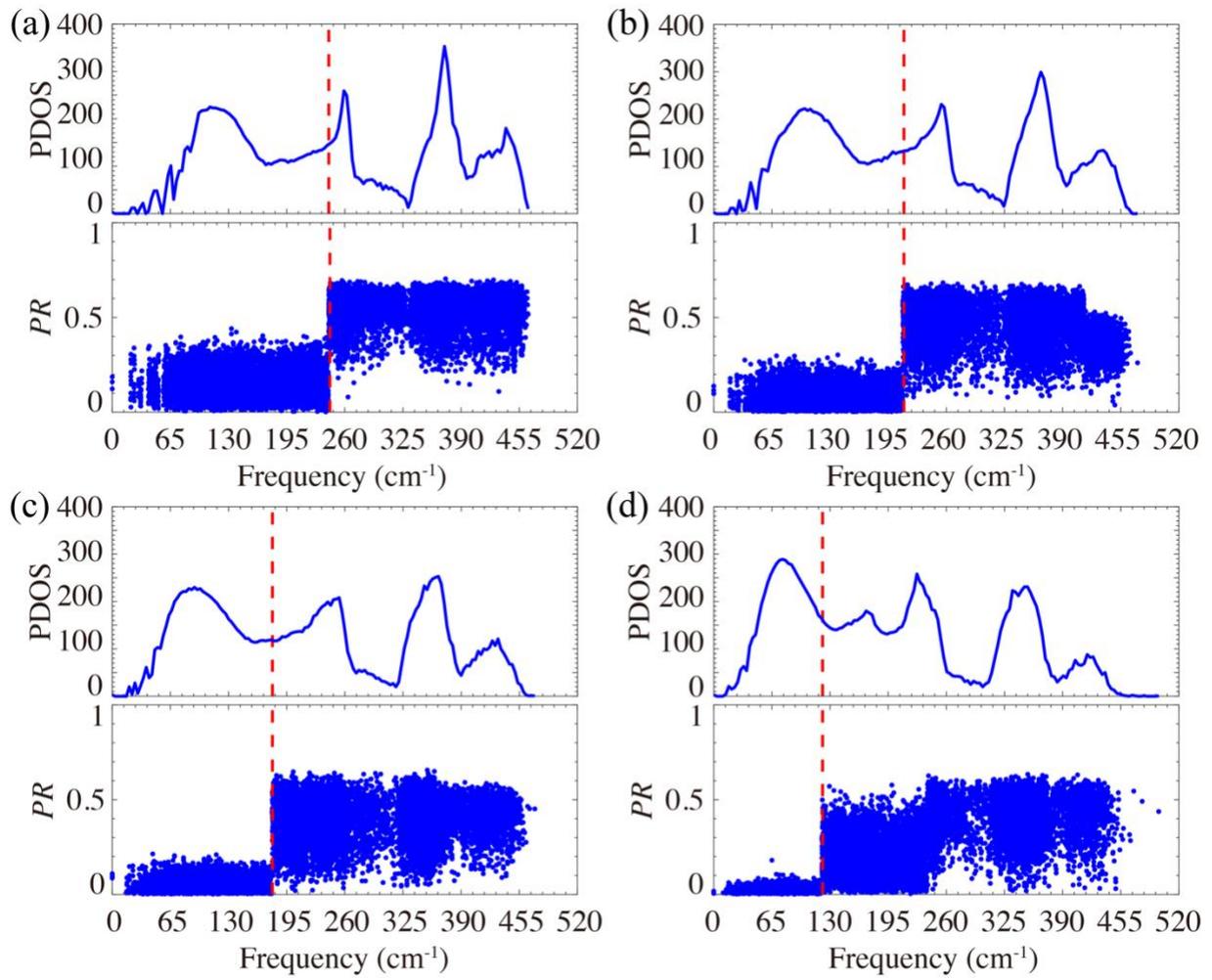

**Figure 6.** Phonon density of states (PDOS) and partition ratio (*PR*) of $Si_yGe_ySn_x$ MEAs with (a) $x = 0$, (b) $x = 1/12$, (c) $x = 1/6$, and (d) $x = 1/3$.

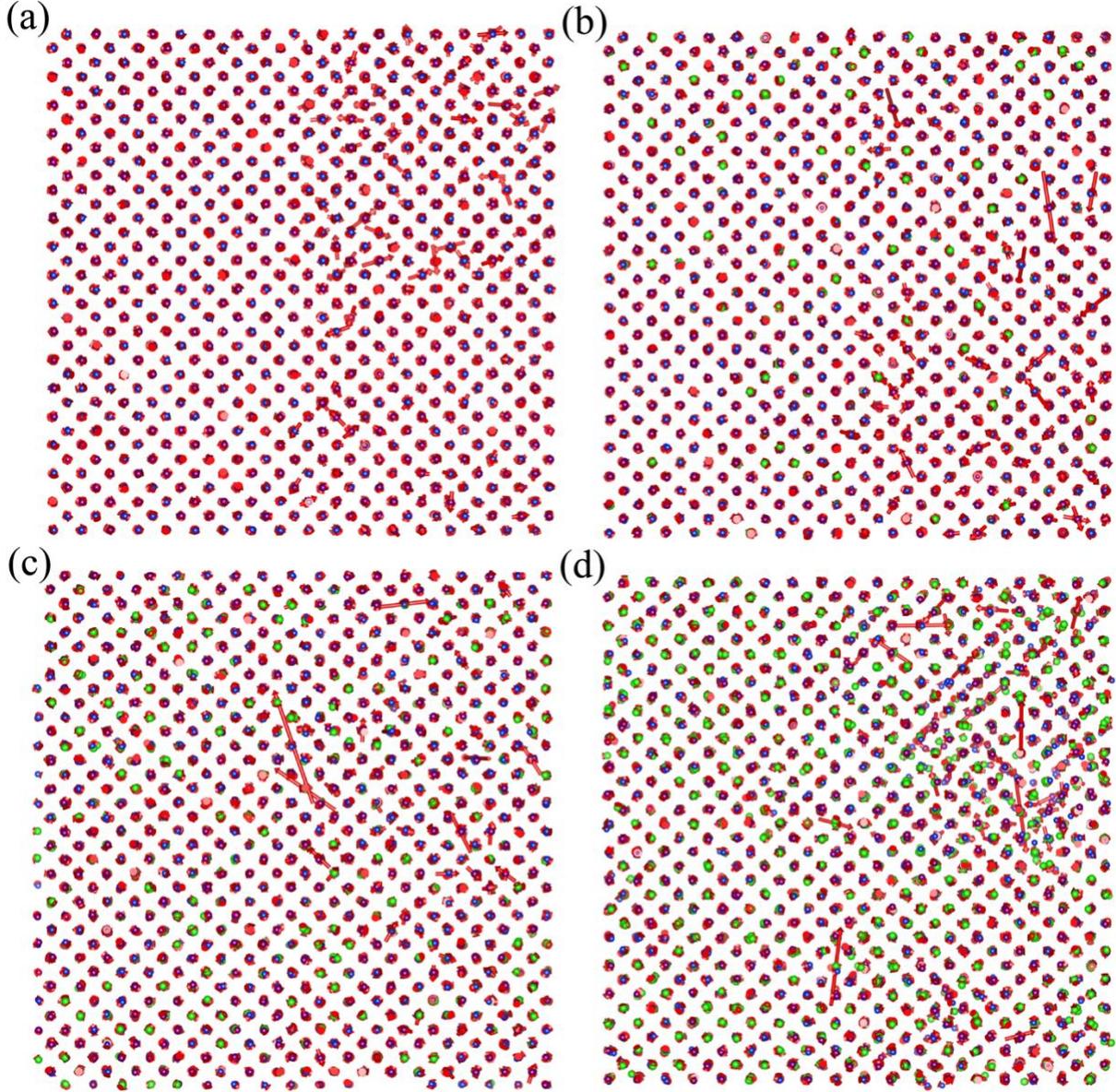

**Figure 7.** Phonon eigenvectors that correspond to the same low partition ratio, 0.022, of $Si_yGe_ySn_x$ MEAs with (a) $x = 0$, (b) $x = 1/12$, (c) $x = 1/6$, and (d) $x = 1/3$. To aid the visualization of the normal modes, we scale the computed eigenvectors by a factor of 20.

In the phenomenal model for understanding the trend in the thermal conductivity of $Si_yGe_ySn_x$ MEAs, we consider phonon scattering by accounting for two effects. The first one is compositional disorder and the second is lattice distortion that has been shown above. Taking these two factors into account, we can approximate the phonon scattering strength from alloy (caused by the compositional disorder) and strain effects (originated from the lattice distortion) as [41, 94],

$$\Gamma = c_i \left[ \left(1 - \frac{M_i}{\overline{M}}\right)^2 + \gamma \left(1 - \frac{a_i}{a_{\text{avg}}}\right)^2 \right], \tag{3}$$

where $M_i$ is the atomic mass of the $i$th element and $\overline{M}$ is the averaged atomic mass dependent on the concentrations $c_i$ of individual elements. Similarly, $a_i$, are the optimized lattice constants of Si, Ge, and $\alpha$-Sn using the empirical Si-Ge-Sn interatomic potential. $a_{\text{avg}}$ is the average lattice constant of a $Si_yGe_ySn_x$ MEA. $\gamma$ is an adjusting parameter taken as 39 based on the previous work on conventional Si-Ge-Sn alloys [94]. The inversed phonon scattering strength $\Gamma_{-1}$ determines the phonon lifetime $\tau$ from phonon scattering. $\tau$ is proportional to $\Gamma_{-1}$ with the proportional factor dependent on phonon group velocities and frequencies and lattice constants of a system. Figure 5(b) shows the variation of $\Gamma_{-1}$ with $x$ in $Si_yGe_ySn_x$ MEAs. We observe the same trend as in the thermal conductivity versus the Sn content, i.e., $\Gamma_{-1}$ decreases as the Sn content increases in the MEAs. This consistent trend implies that the phonon scattering grows increasingly stronger as the compositional disorder increases and thus more thermal energy carriers are scattered rather than transported to conduct heat, leading to the decreased thermal conductivities found in our EMD calculations.

Finally, we attempt to understand the trend the thermal conductivity of $Si_yGe_ySn_x$ MEAs from the perspective of DFT calculations. We first use VASP and Phonopy to compute the second-order force constant and construct a dynamical matrix and obtain the phonon spectra of two artificial Si-Ge-Sn MEAs of $Si_{1/2}Ge_{3/8}Sn_{1/8}$, and $Si_{3/8}Ge_{3/8}Sn_{1/4}$. We see from Figure 8 (a) and (b) that phonon frequencies shown in the phonon spectra are all real, showing that these two MEA systems are dynamically stable. The right plot of each diagram in Figure 8 shows the accumulated thermal conductivity $\Sigma\kappa$ at 300 K and the derivative of the thermal conductivity $d\kappa$ as a function of phonon frequency. We can see that $\Sigma\kappa$ generally increases with phonon frequencies and becomes nearly a constant at high frequencies. Correspondingly, d$\kappa$ increases and peaks and decays to zero. From the saturated $\Sigma\kappa$, we determine the thermal conductivities of Si$_{1/2}$Ge$_{3/8}$Sn$_{1/8}$ and Si$_{3/8}$Ge$_{3/8}$Sn$_{1/4}$ as 41.1 and 19.8 W/(m·K), respectively. We expect these DFT thermal conductivities to be more accurate than those from our EMD simulations. For example, our calculated DFT thermal conductivity of Si (122.9 W/(m·K)) for benchmark is consistent with the experimental value of 125.5 W/(m·K) [91] and with 129.5 W/(m·K) from previous DFT calculations [95]. Both $\Sigma\kappa$ and d$\kappa$ show that the thermal conductivities of the four systems originate mainly from the contributions of phonons with relatively low frequencies. For the integration range from 0 to 100 cm$^{-1}$, the $\Sigma\kappa$ of Si$_{1/2}$Ge$_{3/8}$Sn$_{1/8}$ and Si$_{3/8}$Ge$_{3/8}$Sn$_{1/4}$ reaches 83% and 92%, respectively, of their corresponding total thermal conductivities. Increasing the integration range up to 150 cm$^{-1}$, the $\Sigma\kappa$ of Si$_{1/2}$Ge$_{3/8}$Sn$_{1/8}$ increases to 95%, whereas the $\Sigma\kappa$ of Si$_{3/8}$Ge$_{3/8}$Sn$_{1/4}$ reaches 96%, respectively, of their total thermal conductivities. In other words, comparing to Si$_{1/2}$Ge$_{3/8}$Sn$_{1/8}$, Si$_{3/8}$Ge$_{3/8}$Sn$_{1/4}$ with higher Sn content exhibits a smaller thermal conductivity and the $\Sigma\kappa$ saturates much more rapidly.

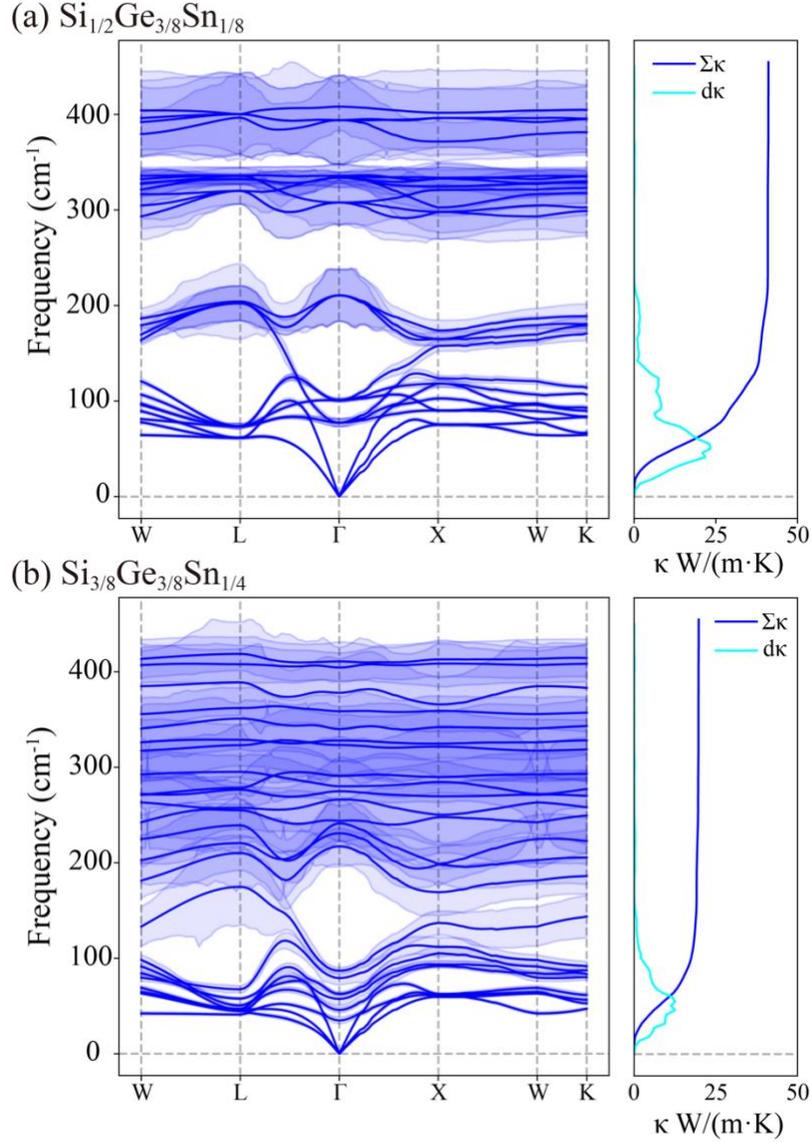

**Figure 8.** Phonon spectrum and linewidth, thermal conductivity derivative dκ, and cumulated thermal conductivity Σκ of two artificial MEAs (a) $Si_{1/2}Ge_{3/8}Sn_{1/8}$, (b) $Si_{3/8}Ge_{3/8}Sn_{1/4}$. To aid the visualization of phonon linewidth, we scale the computed values by a factor of 50.

The anharmonic phonon-phonon interactions play important roles in decreasing the thermal transport of a material. To interrogate the effects of compositional disorder on the anharmonic phonon-phonon interactions and consequently the thermal conductivities of $Si_{1/2}Ge_{3/8}Sn_{1/8}$ and $Si_{3/8}Ge_{3/8}Sn_{1/4}$, we focus on the phonon linewidths caused by anharmonic phonon-phonon interactions. Unlike the linewidth in the phenomenal model that considers the alloy and strain effects, the linewidth in DFT calculations here account for the alloy and Umklapp scattering effects [41, 96]. We can derive the phonon lifetime $\tau_\lambda$ through an inverse relationship with phonon linewidth, $\tau_\lambda = 1/(2\Gamma_\lambda)$. Experimentally, the phonon linewidth $2\Gamma_\lambda$ can be measured via time-resolved Raman spectroscopy, where $2\Gamma_\lambda$ approximately equals to the full width at half maximum of the Raman spectrum peak [97, 98]. Theoretically, anharmonic phonon-phonon interactions are described by the frequency ($\omega$)-dependent phonon self-energy $\Delta(\omega) - i\Gamma(\omega)$ [97]. The real part $\Delta(\omega)$ gives the frequency shift due to the phonon-phonon scattering, whereas the imaginary part $\Gamma(\omega)$ is associated with the probability of phonon decay that can be written as,

$$\Gamma_\lambda(\omega) = \frac{18\pi}{\hbar^2} \sum_{\lambda_1 \lambda_2} |V_{\lambda \lambda_1 \lambda_2}|^2 S(\lambda, \omega). \tag{4}$$

Here, $V_{\lambda \lambda_1 \lambda_2}$ denotes a third-order tensor representing the many-body interactions following the direction for the energy decreasing most rapidly regarding three displacements from the equilibrium positions [99, 100]. $S(\lambda, \omega)$ written below

$$S(\lambda, \omega) = \left\{ (1 + n_{\lambda_1} + n_{\lambda_2}) \delta(\omega_\lambda - \omega_{\lambda_1} - \omega_{\lambda_2}) + (n_{\lambda_1} - n_{\lambda_2}) \left[ \delta(\omega_\lambda + \omega_{\lambda_1} - \omega_{\lambda_2}) - \delta(\omega_\lambda - \omega_{\lambda_1} + \omega_{\lambda_2}) \right] \right\}$$

(5)

result from two types of phonon decay processes: (1) the down-conversion process where the initial phonon is decomposed into two phonons of lower frequencies $\delta(\omega_\lambda - \omega_{\lambda_1} - \omega_{\lambda_2})$, and (2) the up-conversion process where the phonon at non-equilibrium state absorbs another phonon to form a higher-frequency phonon, $\delta(\omega_\lambda + \omega_{\lambda_1} - \omega_{\lambda_2})$ and $\delta(\omega_\lambda - \omega_{\lambda_1} + \omega_{\lambda_2})$ [77, 99, 101]. $n_{\lambda_i}$ in Eq. 8 represents the Bose-Einstein distribution function, i.e., $n_{\lambda_i} = \left\{ \exp\left[\hbar \omega_{\lambda_i} / k_B T\right] - 1 \right\}^{-1}$.

Figure 8 shows the calculated phonon linewidths of Si$_{1/2}$Ge$_{3/8}$Sn$_{1/8}$ and Si$_{3/8}$Ge$_{3/8}$Sn$_{1/4}$ overlapped with the corresponding phonon modes. We observe that large phonon linewidth caused by the anharmonic phonon-phonon interactions in the optical branches suppresses the increase of thermal conductivity. We calculate the average phonon linewidth in different ranges of phonon frequencies. For $Si_{1/2}Ge_{3/8}Sn_{1/8}$, the averaged phonon linewidth of all the phonon modes is 0.563 $cm^{-1}$, whereas the overall averaged phonon linewidth of $Si_{3/8}Ge_{3/8}Sn_{1/4}$ is 0.888 $cm^{-1}$, 58% higher than $Si_{1/2}Ge_{3/8}Sn_{1/8}$. Because the major contribution of thermal conductivity at 300 K results from the relatively lower-frequency phonon modes, we compute the average phonon linewidth for $Si_{1/2}Ge_{3/8}Sn_{1/8}$ and $Si_{3/8}Ge_{3/8}Sn_{1/4}$ using their phonon modes with the frequencies in the range from 0 to 150 $cm^{-1}$. This region is also where the thermal conductivity derivative increases rapidly for all the four systems. Once again, the averaged phonon linewidth of $Si_{3/8}Ge_{3/8}Sn_{1/4}$ (0.150 $cm^{-1}$) is longer than that of $Si_{3/8}Ge_{3/8}Sn_{1/4}$ (0.075 $cm^{-1}$). We therefore conclude that the anharmonic phonon-phonon interactions cause the increase of phonon linewidth in the low-frequency region of $Si_{3/8}Ge_{3/8}Sn_{1/4}$, shortening the phonon lifetime and suppressing the increase of thermal conductivity in the high-frequency region.

With the calculated Seebeck coefficients, electrical conductivities, and thermal conductivities, we compute the figure of merit at room temperature for the 11 $Si_yGe_ySn_x$ MEAs. Because the underestimated EMD thermal conductivities lead to overestimated figure of merit, we scale the computed figure of merit by a factor of 1/2.19 (The denominator comes from the ratio between the DFT and EMD thermal conductivities of Si). Figure 9 shows that, at different chemical potentials, $Si_yGe_ySn_x$ MEAs can achieve figure of merit that is comparable to that of other room-temperature TE materials.

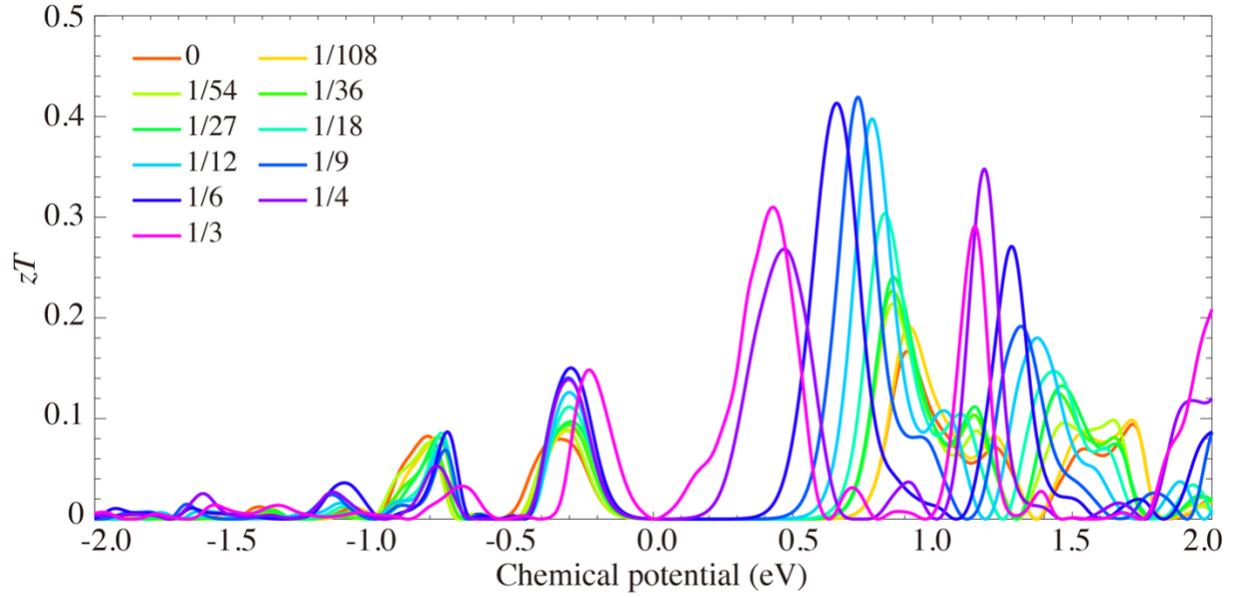

**Figure 9.** Scaling figure of merit of Si$_y$Ge$_y$Sn$_x$ MEAs at different electron chemical potentials. The scaling factor is 1/2.19 due to the underestimated thermal conductivity in the molecular dynamics simulations.

**Conclusion**

We have performed DFT and classical MD calculations to investigate the electrical and thermal transport properties of Si$_y$Ge$_y$Sn$_x$ MEAs. We found that Si$_y$Ge$_y$Sn$_x$ MEAs exhibit a wide range of bandgaps that span from visible to near-infrared in the light spectrum and that Si$_y$Ge$_y$Sn$_x$ MEAs with high Sn content have small electron and hole effective masses. Furthermore, we showed that the electron Anderson localization is not as distinct as phonon Anderson localization. The excellent electrical properties of Si$_y$Ge$_y$Sn$_x$ MEAs make them promising for applications in a variety of electronic devices. Meanwhile, we found Si$_y$Ge$_y$Sn$_x$ MEAs show low thermal conductivities, owing to the phonon Anderson localization and strong anharmonic phonon-phonon interactions caused by the lattice distortion and compositional disorder. The high electrical conductivities and low thermal conductivities endow Si$_y$Ge$_y$Sn$_x$ MEAs with great potential for high TE performance at room temperature. Our prediction calls for future experimental verification and also for simulation models to obtain more accurate thermal transport properties for these MEAs.

**Acknowledgements**

We thank John Kouvetakis and José Menendéz at Arizona State University (ASU) for helpful discussion. We also thank Yongjin Lee for sharing the Si-Ge-Sn interatomic potential used in this work. We acknowledge the start-up funds from ASU. This research used computational resources of the Texas Advanced Computing Center under Contracts No.TG-DMR170070 and the Agave cluster at ASU.